\documentstyle[twocolumn,floats,epsf,prb,aps]{revtex}
\newcommand{\beq}{\begin{equation}}
\newcommand{\eeq}{\end{equation}}
\newcommand{\bea}{\begin{eqnarray}}
\newcommand{\eea}{\end{eqnarray}}
\newcommand{\bi}{\begin{itemize}}
\newcommand{\ei}{\end{itemize}}
\newcommand{\be}{\begin{enumerate}}
\newcommand{\ee}{\end{enumerate}}
\newcommand{\bd}{\begin{description}}
\newcommand{\ed}{\end{description}}
\newcommand{\bfig}{\begin{figure}}
\newcommand{\efig}{\end{figure}}

\newcommand{\phrb}{Phys. Rev. B } 
 
\newcommand{\phrl}{Phys. Rev. Lett. }

\begin{document} 
\twocolumn[\hsize\textwidth\columnwidth\hsize
\csname @twocolumnfalse\endcsname
\draft
 
\title{Monte Carlo Study of the Finite Size Effects on the Magnetization
of Maghemite Small Particles} 
\author{\`Oscar Iglesias, Am\'{\i}lcar Labarta, and F\`elix Ritort} 
\address{Departament de F\'{\i}sica Fonamental, 
Universitat de Barcelona,  
Diagonal 647, 08028 Barcelona, Spain} 
\maketitle   
\begin{center} 
(Received   
\end{center}
\begin{abstract} 
Monte Carlo simulations of a model for $\gamma$-Fe$_2$O$_3$ (maghemite) 
single particle of spherical shape are presented aiming at the elucidation
the specific role played by the finite size and the surface on 
the anomalous magnetic behaviour observed in small particle 
systems at low temperature. 
The influence of the finite-size effects on the equilibrium properties of 
extensive magnitudes, field 
coolings and hysteresis loops is studied an compared to the results for 
periodic boundaries.  
It is shown that for the smallest sizes the thermal demagnetization of the 
surface completely dominates the magnetization while the behaviour of the 
core is similar to that of the periodic boundary case,
independently of $D$. The change in shape of the hysteresis loops with 
$D$ demonstrates that the reversal mode is strongly influenced by 
the presence of broken links and disorder at the surface.
\end{abstract} 
\date{\today}
]
$\gamma$-Fe$_2$O$_3$ (maghemite) and other magnetic oxides in the form of 
nanometric particles display anomalous magnetic properties at low temperatures.
Experiments have shown that the hysteresis 
loops display high closure fields and do not saturate 
\cite{Kodama97,Garcia99,Martinez98} even at fields of 
the order of 50 T. 
Low magnetization as compared to bulk, shifted loops after field cooling 
and irreversibilities between the field cooling and zero field cooling processes 
even at high fields are also observed \cite{Garcia99,Martinez98}. 
Moreover, the existence of aging phenomena \cite{Jonsson95} in the time-dependence 
of the magnetization, indicates that there must be some kind of freezing 
leading to a complex hierarchy of energy levels. 
Whether these phenomena can be ascribed to intrinsic properties 
of the particle itself (spin-glass (SG) state of the surface which creates an 
exchange field on the core of the particle), or they are due to a collective 
behaviour induced by interparticle interactions \cite{Batlle97,Dormann98,Morup94}, 
has been the object of controversy in recent years and up to the moment there is 
no model giving a clear-cut explanation of this 
phenomenology, although simulation results for general small particle 
systems \cite{Trohidou} and, in particular, for maghemite
\cite{Kach00} have been recently published. 
In order to contribute to elucidate this controversy we present the results of 
a Monte Carlo (MC) simulation of a single spherical particle which aims at clarifying 
what is the specific role of the finite size and surface on the magnetic 
properties of the particle, disregarding the interparticle interactions effects.
 
Maghemite is a ferrimagnetic spinel in which the magnetic Fe$^{3+}$ ions with spin 
$5/2$ are disposed in two sublattices with different oxigen coordination 
(8 tetrahedric (T) and 16 octahedric (O) sites per unit cell). 
In our model, the Fe$^{3+}$ magnetic ions are represented by Ising spins 
$S_i^{\alpha}=\pm 1$ which allows to reproduce a case with strong uniaxial 
anisotropy while keeping computational efforts within reasonable limits.
The spins interact via antiferromagnetic exchange interactions with the nearest 
neighbours on both sublattices and with an external magnetic field $H$.
In the simulation, we have used the reduced field $h=\mu H/k_{B}$ in temperature
units, where $\mu$ is the magnetic moment of the Fe$^{3+}$ ions.
The values of the nearest neighbour exchange constants for maghemite are \cite{Kach00} 
$J_{TT}=-21\ $K, $J_{OO}= -8.6\ $K, $J_{TO}= -28.1\ $K. 
We have used periodic boundary (PB) conditions to simulate the bulk properties 
and free boundaries (FB) for a spherically shaped particle with $D$ unit cells 
in diameter when studying finite size effects. 
In the latter case, two different regions are 
distinguished: the surface formed by the outermost unit cells 
and an internal core. The size of the studied 
particles ranges from $D= 3$ to $14$ corresponding to real particle diameters 
from 2.49 to 11.62 nm. 
 
We start by studying the effect of FB conditions and  
finite-size effects on the equilibrium properties in zero magnetic field. 
The simulations have been performed using the standard Metropolis algorithm.  
Starting from a very high temperature ($T= 200\ $K) and an initially disordered state  
with spins randomly oriented, the system was cooled down at a constant  
temperature step $\delta T= -2\ $K. After discarding the first 1000 MC steps, 
the averages of the  
thermodynamic quantities were computed at each temperature during a number  
of MC steps ranging from 10000 to 50000 depending on  
the system size.  
 
In Fig. \ref{MMM01_M(T)_fig}, we compare the thermal dependence of the 
magnetization for 
spherical particles of different diameters $D$ with the corresponding 
results for a system of size $N=14$ and PB (uppermost curve).
A second order transition from paramagnetic to ferrimagnetic order  
signaled by a sharp peak at $T_c (D)$ in the susceptibility  
(see the Inset in Fig. \ref{MMM01_M(T)_fig}) is clearly observed.  
Finite-size effects on both the magnetization and the susceptibility are very 
important even for $D$'s as large as 14 in the FB case, with
$T_c (D)$ increasing as $D$ increases and tending to the value for PB
(which varies from 122 to 126 K when increasing $N$ from 3 to 14). 
The main feature in Fig. \ref{MMM01_M(T)_fig} is the reduction of the magnetization  
$M$ with respect to the PB case (dashed line) due to the lower  
coordination of the spins at the surface, which hinders perfect 
ferrimagnetic order  at finite temperatures.  
It is worthwhile to note that for all the studied diameters there is a 
temperature range, in which the demagnetization process of $M$ is linear, 
this range being wider as the particle size decreases. 
In this linear regime the particle demagnetization becomes dominated by the 
surface effects, being the core and surface behaviours strongly correlated. 
Linear demagnetization is indicative of the effective 3D-2D dimensional 
reduction of the surface spins and has previously been observed in thin film 
systems \cite{Martinez92}. 

\bfig[t] 
\centering 
\leavevmode  
\epsfxsize=8 cm  
\epsfbox{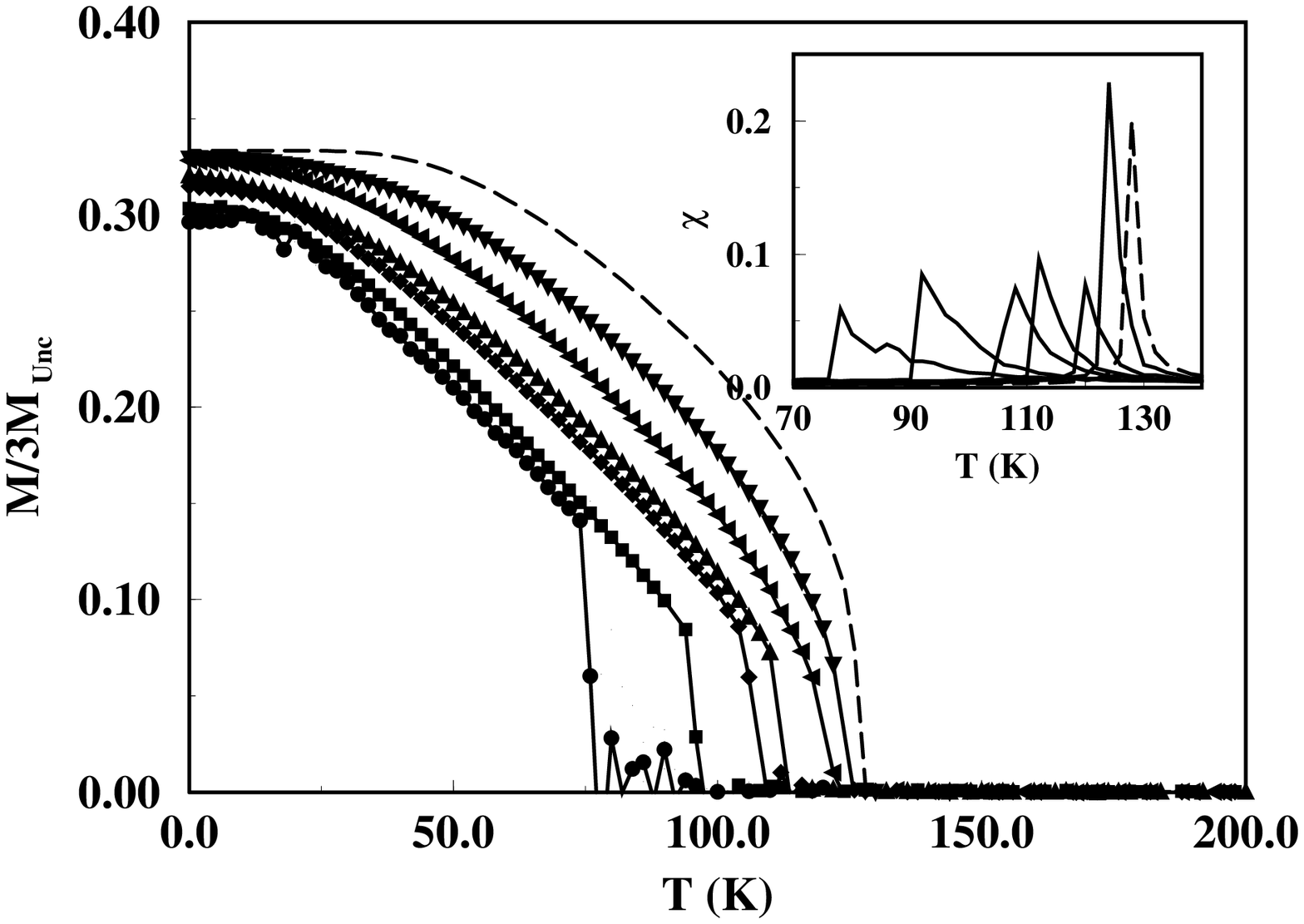}  
\caption{
Thermal dependence of the magnetization $M$.
The results for particle diameters $D= 3, 4, 5, 6, 8, 14$ 
(from the lowermost curve in circles) and PB conditions  
$N= 14$ (uppermost curve) are shown. Inset: Thermal dependence of the susceptibility 
for the same cases (the PB case is drawn in dashed lines). 
$M_{Unc}$ is the ratio of the difference of O and T spins to the total number 
of spins.  
}  
\label{MMM01_M(T)_fig} 
\efig 
\bfig[t] 
\centering 
\leavevmode  
\epsfxsize=8 cm  
\epsfbox{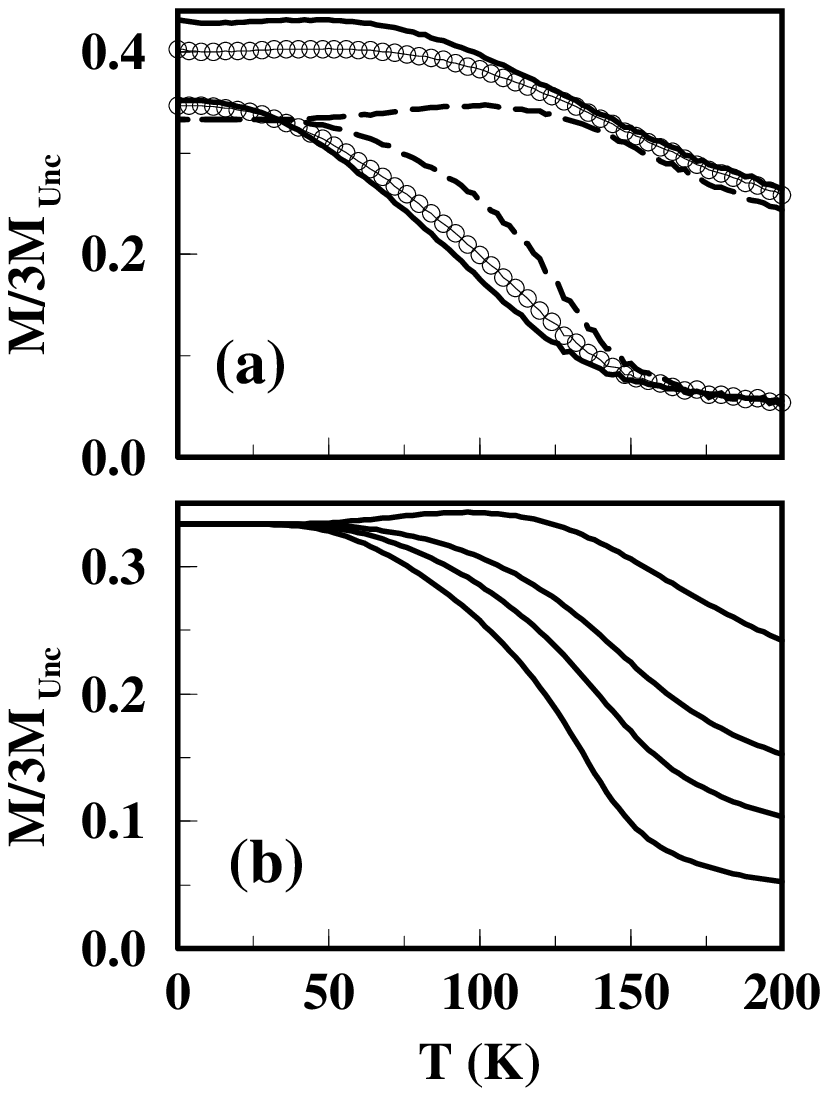}  
\caption{ 
Thermal dependence of $M$ after cooling under $h_{FC}$. 
(a) corresponds to a spherical particle with $D= 6$. 
The results for two cooling fields $h_{FC}= 20\ $K (lower curve) and  
$h_{FC}= 100\ $K (upper curve) are shown. 
The contributions of the surface (thick lines) and the core (dashed lines) 
to the total magnetization (circles) have been plotted separatedly.  
(b) shows $M$ for an $N= 8$ system and PB after cooling in $h_{FC}= 20, 40, 80, 100\ $K 
(from lower to uppermost curves). (See Fig. 1 for the definition of $M_{Unc}$). 
}  
\label{MMM01_FC_fig} 
\efig 
Deeper insight on the magnetic ordering of this system can be gained by studying 
the thermal dependence of the equilibrium magnetization in a  
magnetic field. Several such curves are presented in Fig. \ref{MMM01_FC_fig}.  
They have been obtained by the same cooling procedure as described previously  
in presence of different fields $h_{FC}$. 
In this figure, the surface (continuous lines) and the core 
(dashed lines) contributions to the total magnetization have been distinguished.
For PB all the curves tend to the ferrimagnetic order value (i.e. $M=1/3$). 
A maximum appears at high enough cooling fields $h_{FC}= 100\ $K which 
is due to the competition between the ferromagnetic 
alignment induced by the field and the spontaneous ferrimagnetic order 
(as the temperature is reduced the strength of the field is not enough 
as to reverse the spins into the field direction).  
However, for particles of finite size the curves at different fields do 
not converge to the ferrimagnetic  
value at low $T$, reaching higher values of the magnetization at $T= 0$ 
the higher the cooling field (see lines with circles in Fig. 
\ref{MMM01_FC_fig}a). 
The total magnetization for small particles is  
completely dominated by the surface contribution 
and this is the reason why the ferrimagnetic order is  
less perfect at these small sizes and the magnetic field can easily  
magnetize the system.  
However, the behaviour of the core is still  
very similar to that of the case with PB, although its contribution  
to $M$ is small.  
At low fields, the surface is in a more disordered state than the core:  
its magnetization lies below $M$ at temperatures for which the thermal  
energy dominates the Zeeman energy of the field.
In contrast, a high field is able to magnetize the surface easier than the  
core due to the fact that the broken links at the surface worsen the  
ferrimagnetic order while the core spins align towards the field  
direction in a more coherent way.

\bfig[t] 
\centering 
\epsfxsize=8 cm  
\leavevmode  
\epsfbox{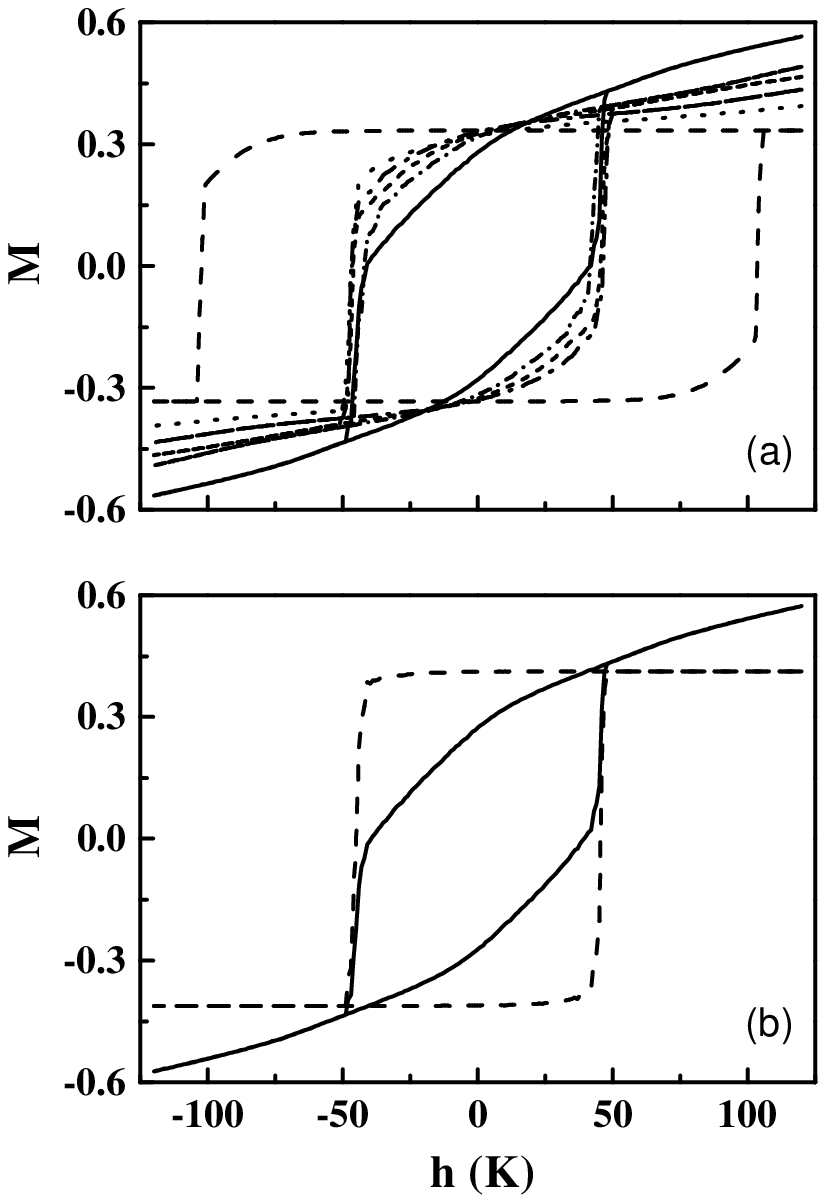}  
\caption{(a) Hysteresis loops for particles of diameters $D=3, 4, 6, 8, 10$
(from inner to outermost curves) and PB (long-dashed lines) at $T= 20\ $K. 
(b) Surface (continuous line) and core (dashed line) 
contributions for a particle of diameter $D= 3$ at 
$T= 20\ $K.}  
\label{MMM01_HIST_fig} 
\efig 
In Fig. \ref{MMM01_HIST_fig}a, we show the hysteresis loops 
for several particle diameters at $20\ $K. 
The loops have been computed by starting  
from a demagnetized state at $h= 0$ (the first magnetization
curve is not shown in this figure).  
The results have been averaged over several ($\approx 10$) independent runs 
starting with different random number seeds. 
First of all, let us note that the saturation field and high-field 
susceptibility increase as the particle size is reduced,  
since these quantities are mainly associated to the  
progressive alignment of the surface spins towards the field direction.  
As a consequence, the cycles of the smallest particles are similar to
those observed in disordered systems, increasing their squareness 
(uniform rotation of $M$) with the size.
In fact, by plotting separatedly the contributions of the core and surface 
to the total magnetization (see Fig. \ref{MMM01_HIST_fig}b), we see that the  
loop of the core is almost perfectly squared independently of the  
particle size, indicating a coherent reversal of its magnetization, 
while the loop of the surface reveals a progressive reversal of $M$, 
which is a typical feature of a disordered or frustrated system. 
Nonetheless, the coercive field of the core is slightly higher but very 
similar to the one of the surface indicating that the reversal of the surface 
spins triggers the reversal of the core. 
We have shown that the existence of lower coordination at the surface
of the particle hinders perfect ferrimagnetic order, increasing the
magnetic disorder at the surface layer as the size of the particle decreases. 
However, the magnetic frustration associated to the competition between 
intra- and inter-sublattice antiferromagnetic interactions,
finite-size and surface effects are not enough as to produce a 
SG layer contrary to
the experimental observation of some authors \cite{Kodama97,Martinez98}. 
This indicates that other ingredients (i.e., enhanced anisotropy at the
surface) should be included in the model to account for the SG layer.
Our model shows that magnetic disorder at the surface simply facilitates 
the thermal demagnetization of the particle and also 
increases the magnetization at moderate fields, since surface disorder 
diminishes ferrimagntic correlations within the particle. 
\section*{Acknowledgements}
We acknowledge CESCA and CEPBA under coordination of 
C$^4$ for the computer facilities. 
This work has been supported by 
SEEUID through project MAT2000-0858 and CIRIT under project 2000SGR00025.


\begin{thebibliography}{99}

\bibitem{Kodama97}
R. H. Kodama, S. A. Makhlouf, and A. E. Berkowitz,
{\phrl} {\bf 79}, 1393 (1997).

\bibitem{Garcia99}
M. Garc\'{\i}a del Muro, X. Batlle, and A. Labarta,
{\phrb} {\bf 59}, 13584 (1999).

\bibitem{Martinez98}
B. Mart\'{\i}nez, X. Obradors, Ll. Balcells, A. Rouanet, and C. Monty ,
{\phrl} {\bf 80}, 181 (1998).

\bibitem{Jonsson95}
T. Jonsson, J. Mattsson, C. Djurberg, F. A. Khan, P. Nordblad, and P. Svedlindh, 
{\phrb} {\bf 75}, 4138 (1995);
T. Jonsson, P. Nordblad, and P. Svedlindh,
{ibid.} {\bf 57}, 497 (1998).

\bibitem{Batlle97}
X. Batlle,  M. Garc\'{\i}a del Muro, and A. Labarta,
{\phrb} {\bf 55}, 6440 (1997).

\bibitem{Dormann98}
J.L. Dormann, R. Cherkaoui, L. Spinu, M. Nogu\'es, F. Lucari, F.
D'Orazio, D. Fiorani, A. Garc\'{\i}a, E. Tronc, and J.P. Jolivet,
{J. Magn. Magn. Mat.} {\bf 187}, L139 (1998).

\bibitem{Morup94}
S. Morup and E. Tronc,
{\phrl} {\bf 72}, 3278 (1994).

\bibitem{Trohidou}
K. N. Trohidou and J. A. Blackman,
{\phrb} {\bf 41}, 9345 (1990);
D. A. Dimitrov and G. M. Wysin,
{ibid.} {\bf 50}, 3077 (1994);
{ibid.} {\bf 51}, 11947 (1995).

\bibitem{Kach00}
H. Kachkachi, A. Ezzir, M. Nogu\'es, and E. Tronc,
{Eur. Phys. J. B} {\bf 14}, 681 (2000);
H. Kachkachi, M. Nogu\'es, E. Tronc, and D. A. Garanin,
{J. Magn. Magn. Mat.} {\bf 221}, 158 (2000).;
R. H. Kodama and A. E. Berkowitz,
{\phrb} {\bf 59}, 6321 (1999).

\bibitem{Martinez92}
B. Mart\'{\i}nez and R. E. Camley,
{J. Phys. CM} {\bf 4}, 5001 (1992);
G. Xiao and C. L. Chien,
{J. Appl. Phys.} {\bf 61}, 4313 (1987);
A. Corciovei, {Phys. Rev.} {\bf 130}, 2223 (1963).

\end{thebibliography}
\end{document}